\documentclass[conference]{IEEEtran}

\usepackage{amsmath,amssymb,amsfonts}

\usepackage{amsthm}

\usepackage{graphicx}
\usepackage{booktabs}
\usepackage{array}
\usepackage{multirow}
\usepackage{algorithm}
\usepackage{algpseudocode}
\usepackage{cite}
\usepackage[hyphens]{url}

\theoremstyle{definition}
\newtheorem{definition}{Definition}

\theoremstyle{plain}
\newtheorem{proposition}{Proposition}

\algrenewcommand\algorithmicrequire{\textbf{Input:}}
\algrenewcommand\algorithmicensure{\textbf{Output:}}

\usepackage{xcolor}
\definecolor{SJGLink}{HTML}{214B75}
\usepackage[colorlinks=true,linkcolor=SJGLink,citecolor=SJGLink,urlcolor=SJGLink]{hyperref}

\usepackage{flushend}
\usepackage{needspace}
\newcommand{\searchbreadth}{\mathit{ef}}
\begin{document}
\bstctlcite{SJG-author-format}

\title{SJG: A Skyline Joint Graph for Multi-Component Approximate Nearest Neighbor Search}

\author{\IEEEauthorblockN{Linyun Ma}
\IEEEauthorblockA{\textit{College of Computer Science and Software Engineering}\\
\textit{Shenzhen University}\\
Shenzhen, China\\
\href{https://orcid.org/0009-0006-0597-2775}{0009-0006-0597-2775}}
\and
\IEEEauthorblockN{Rui Mao}
\IEEEauthorblockA{\textit{College of Computer Science and Software Engineering}\\
\textit{Shenzhen University}\\
Shenzhen, China\\
\href{https://orcid.org/0000-0002-3645-5520}{0000-0002-3645-5520}}
}

\maketitle

\begin{abstract}
Graph-based indexes enable efficient approximate nearest neighbor search at scale by linking similar objects. In applications such as product retrieval and rental listing search, similarity depends on several components whose relative importance varies across queries. The challenge is that the index must select neighbors during construction, but the query weights that determine which objects are closest are not yet known. A straightforward approach is to index each component separately. However, merging the results of these searches can miss the nearest neighbor under the weighted distance. A joint graph allows a search to follow links across components, helping to avoid such misses, but selecting neighbors separately within each component can leave redundant edges, and keeping component hierarchies synchronized limits parallel construction. To address these limitations, we propose the Skyline Joint Graph (SJG). Although query weights can change the ranking of candidate neighbors, the dominance relation among them is independent of those weights. SJG exploits this property by pooling candidates across components and retaining the Skyline for each node, thereby enabling efficient approximate nearest neighbor search under different query weights while supporting parallel graph construction. Extensive experiments on six real-world datasets demonstrate that SJG achieves up to 27.9$\times$ the query throughput of baseline methods at 90\% Recall@10, while accelerating single-threaded index construction by up to 201$\times$. Its index requires at least 36\% less space than the smallest baseline index.
\end{abstract}

\suppressfloats[t] 
\begin{figure}[t]
\centering
\includegraphics[width=\columnwidth]{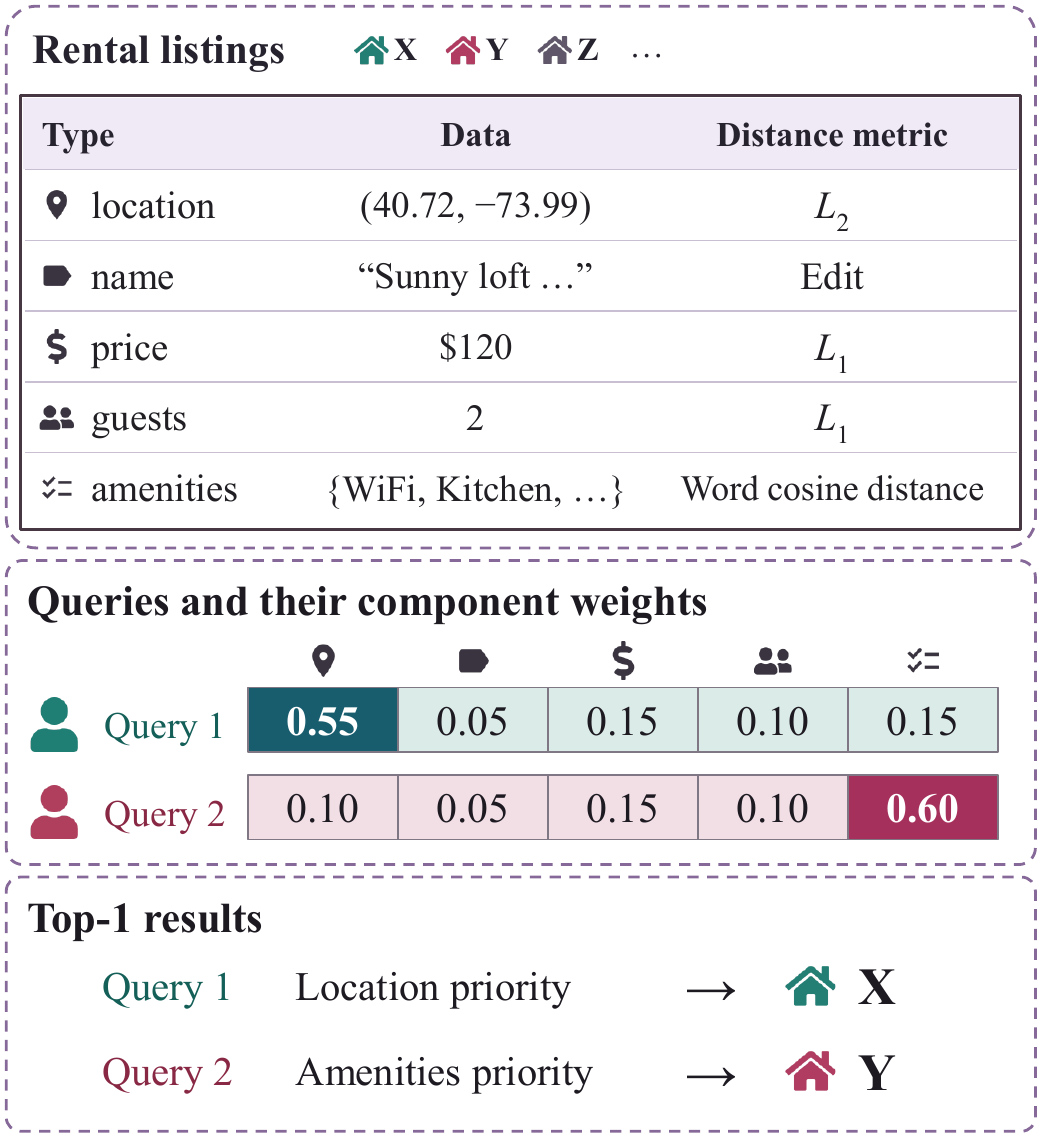}
\caption{Rental listing retrieval under different query weights.}
\label{fig:overview}

\end{figure}

\section{Introduction}\label{sec:intro} 

Similarity search is a fundamental operation in modern data systems. Recommendation engines, retrieval systems, and vector databases all need to retrieve objects that rank highly in similarity to a given query object. Over the past decade, approximate nearest neighbor search (ANNS) has made this task tractable at scale, and graph-based indexes such as HNSW~\cite{HNSW}, NSG~\cite{NSG}, and Vamana~\cite{DiskANN} are widely used in large-scale ANNS~\cite{Azizi25Graph,HSCG}. By organizing objects into a proximity graph, they achieve high recall while visiting only a small fraction of the collection~\cite{Falcon25}. During construction, each object first gathers a set of candidate neighbors and then retains a compact, well-connected subset. Most graph-based ANNS indexes construct and traverse the graph using the same fixed distance function.

As data becomes increasingly diverse, however, objects are no longer well described by a single vector. For example, a rental listing carries a location, a name, a price, a guest capacity, and a set of amenities; a product carries an image, a description, and a price; a user profile carries demographics, behavior, and text. Each component is naturally compared by its own distance function, and different queries attach different importance to these components. As \hyperref[fig:overview]{Fig.~\ref*{fig:overview}} shows, one user may prioritize location in rental listing search, while another may prioritize amenities. The same collection can return different nearest neighbors under different weight assignments. Vector databases already support this kind of retrieval. Milvus~\cite{Milvus}, for example, lets applications search across multiple vector fields and specify how much each field contributes to the ranking. For graph-based indexing, however, supporting such queries introduces a construction challenge. The graph must be built before the query weights are known, even though these weights can change the relative ranking of candidate neighbors.

\begin{figure}[t]
\centering
\includegraphics[width=\columnwidth]{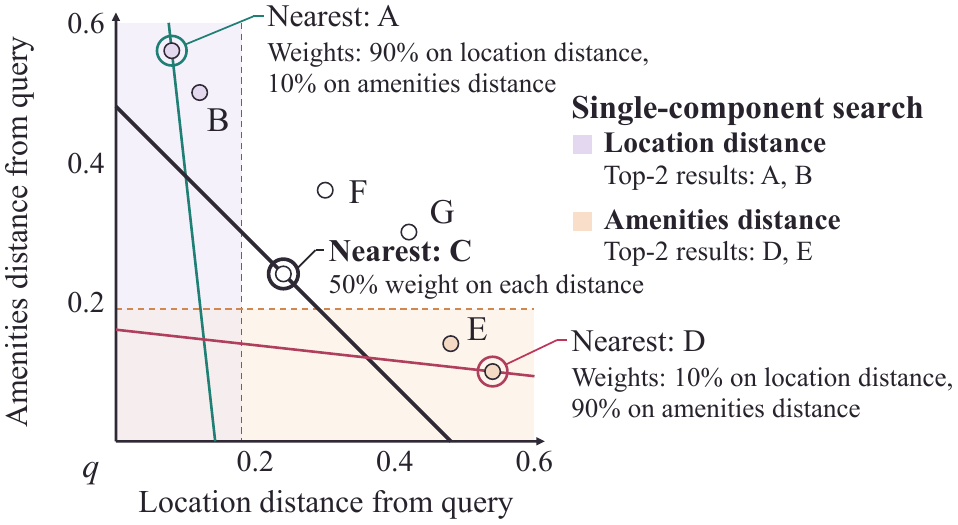}
\caption{Component-wise top-2 results can miss the weighted nearest neighbor.}
\label{fig:component-miss}
\end{figure}

Existing methods address this challenge in different ways. Composite methods build one graph per component and merge the results during query processing. Since each graph selects neighbors using only its own component distance, the query weights are not needed during construction. During search, the system queries the component indexes and combines the retrieved candidates according to the specified weights. Milvus~\cite{Milvus}, for example, supports this pattern for multiple vector fields. Uemura et al.~\cite{Uemura24} use intermediate search results to select a starting node for the next component graph, reducing redundant node visits.

However, simply taking the top candidates from each component and merging them can miss the nearest neighbor under the requested weight combination. As illustrated in \hyperref[fig:component-miss]{Fig.~\ref*{fig:component-miss}}, object C lies outside the top two under both location distance and amenities distance, yet it becomes the nearest neighbor when the two components receive equal weight. Even if both component searches return their exact top two objects, their union excludes C, and weighted reranking cannot recover it. Cross-graph guidance can reduce such misses when components are correlated, but it does not eliminate them in general. Although longer candidate lists may include such objects, they also increase the cost of traversal and reranking. Joint graph indexing takes a different approach, allowing queries to explore neighbor relationships across components and discover objects that separate component searches may miss.

The Hierarchical Joint Graph (HJG)~\cite{HJG} implements this approach by building a hierarchical graph for each component and synchronizing the levels of corresponding nodes across these graphs. However, this joint organization introduces two limitations. First, HJG selects edges separately within each component, which can leave redundant edges in the joint graph and increase both index size and search cost. Second, synchronizing the hierarchies creates dependencies among graph updates, making parallel construction difficult and increasing construction cost.

Taken together, these limitations reveal two related but distinct challenges in constructing a joint graph for queries with unknown weights. The first is how to select effective neighbors when the relative importance of different components is not known in advance. The second is how to facilitate parallel joint graph construction without the dependencies introduced by synchronized component hierarchies.

\noindent\textbf{Challenge I: How to select effective neighbors without knowing the query weights}

Directly combining neighbors selected within individual components introduces substantial redundancy, inflating graph size and search cost. Conventional graph indexes control graph density through neighbor selection, typically using a fixed distance function to determine which edges to retain. With per-query weights, however, the usefulness of a neighbor may change across different distance combinations. Selecting neighbors according to any particular weight assignment may therefore compromise the graph's effectiveness for other queries, while retaining all candidates leads to excessive redundancy. We thus need a strategy that neither depends on a fixed weight assignment nor requires maintaining weight-dependent edge information. To address this challenge, we propose the Skyline Joint Graph (SJG), which uses weight-independent Skyline pruning to remove dominated candidates while preserving promising neighbors across different query weights.

\noindent\textbf{Challenge II: How to facilitate parallel joint graph construction}

Effective neighbor selection is only half of the problem. The construction process must also avoid costly dependencies among graph updates. In HJG~\cite{HJG}, each node must keep the same level across all component graphs. During insertion, a larger neighborhood may promote a node to a higher level, which in turn requires updates across the graphs. These updates can affect neighboring nodes and trigger further promotions, creating cascading dependencies. As a result, insertions cannot proceed independently, making parallel construction difficult and expensive. SJG avoids these dependencies by separating candidate acquisition from joint neighbor selection. The component graphs are built independently, and each node then selects its final neighbors without modifying the adjacency of others. This separation enables parallel construction without hierarchical synchronization or cascading updates.

The main contributions of this paper are summarized as follows.

\begin{itemize}
\item[\textbullet] \textbf{Effective graph indexing with joint neighbor selection.} We propose the Skyline Joint Graph (SJG), which integrates candidates from multiple component-specific graphs and performs joint neighbor selection through Skyline-based pruning. This weight-independent selection reduces redundant edges and yields a compact graph that supports efficient weighted ANNS. The resulting adjacency preserves the minimum weighted distance within each candidate set for every nonnegative weighting.

\item[\textbullet] \textbf{Efficient and parallel graph construction.} We develop a two-stage construction strategy that separates component-specific candidate acquisition from joint neighbor selection. Candidate graphs are built independently, and each node selects its final neighbors from its own candidate set. This design removes hierarchical synchronization and cascading updates, enabling efficient parallel construction.

\item[\textbullet] \textbf{Extensive experimental evaluation.} We evaluate SJG under seven weight settings on each of six real-world datasets and examine its scalability up to 7.75 million objects. SJG outperforms the fastest composite baseline in 36 of the 42 settings, with a median speedup of 2.05$\times$ and a maximum of 27.9$\times$. It builds up to 201$\times$ faster than HJG and produces graph indexes at least 36\% smaller than all evaluated baselines.

\end{itemize}

The rest of this paper is organized as follows. \hyperref[sec:related]{Section~\ref*{sec:related}} reviews related work. \hyperref[sec:problem]{Section~\ref*{sec:problem}} defines multi-component similarity search. \hyperref[sec:method]{Section~\ref*{sec:method}} presents the SJG index and its construction and search algorithms. \hyperref[sec:exp]{Section~\ref*{sec:exp}} reports the experimental results. \hyperref[sec:conclusion]{Section~\ref*{sec:conclusion}} concludes the paper.

\section{Related Work}\label{sec:related}

This section reviews three areas of related work: graph-based ANNS on single-component data, multi-component exact retrieval, and multi-component approximate retrieval.

\subsection{Graph-Based ANNS on Single-Component Data}

Graph-based ANNS indexes represent objects as vertices and search for near neighbors by following edges toward the query~\cite{GraphSurvey21}. A natural starting point is a $k$-nearest neighbor graph, in which each object is connected to its closest neighbors. NN-Descent~\cite{NN-Descent} constructs an approximation to this graph by iteratively comparing objects that occur in one another's neighborhoods. Such a graph captures local proximity, but several neighbors of a vertex may lead a search into much the same region. The choice of edges therefore matters even when good candidate neighbors are already available.

HNSW~\cite{HNSW} organizes its edges across multiple layers. Search uses the sparse upper layers to move toward the query and then descends to a denser neighborhood to refine the result. NSG~\cite{NSG} builds a single-layer graph from an approximate $k$-nearest neighbor graph. For each vertex, it gathers candidates through graph search, prunes redundant edges, and then checks that every vertex is reachable from the entry point. Vamana~\cite{DiskANN} also searches the graph to obtain candidates, but applies a tunable pruning rule that controls when a selected neighbor can rule out another candidate. Other work addresses the cost of building these indexes through candidate edge pruning~\cite{YangPG25}, graph refinement~\cite{MIRAGE25}, and merging existing indexes~\cite{FGIM26}.

Graph search can accommodate a variety of distance functions, as illustrated by NMSLIB~\cite{NMSLIB}. In the methods above, however, the distance function used to select neighbors is fixed during construction. In multi-component retrieval, two candidates can exchange their order of proximity to the same vertex when the query weights change. An edge pruned under one weighting may therefore lead to a neighbor preferred under another.

\subsection{Multi-Component Exact Retrieval}

Exact retrieval over multi-component data primarily relies on distance bounds to exclude objects that cannot be query results. Since different queries may assign different weights to individual components, indexes must retain component-specific distance information from which bounds can be derived for the given query weights.

Existing studies have explored different ways to organize this information. Unified indexes, such as M$^2$-tree~\cite{M2-tree} and M$^3$-tree~\cite{M3-tree,Bustos12}, organize multi-component information within a common structure, with M$^3$-tree retaining component-specific covering information to support queries under varying weights. Another approach organizes distance information through component-specific indexes or reference objects, as explored by PM-Tree and RR*-Tree~\cite{Franzke16}. DESIRE~\cite{DESIRE} uses a cluster-based index that supports query filtering through clusters and their associated distance information. Beyond single-machine indexing, OneDB~\cite{OneDB} investigates multi-metric similarity search in distributed environments.

Unlike these exact methods, which rely on distance bounds for candidate pruning, graph-based approximate retrieval guides search through pre-established neighbor relationships.

\subsection{Multi-Component Approximate Retrieval}

Existing multi-component approximate methods differ in how component information is incorporated into candidate discovery, graph construction, and query traversal.

One approach maintains separate indexes for individual components and combines their search results. Milvus~\cite{Milvus} supports searching different vector fields independently, followed by weighted aggregation and reranking. Since reranking operates only on returned candidates, this approach may miss objects that rank poorly within individual components but are close under the weighted distance. To exploit information across component searches, Uemura et al.~\cite{Uemura24} investigate coordinated graph search, reusing previously retrieved candidates to select entry points for subsequent component searches and evaluating candidates using the weighted distance.

Component information can also be fused during graph construction rather than coordinated only during search. MUST~\cite{MUST} learns modality weights to construct fused representations and builds a proximity graph using similarities between weighted concatenated representations for target-modality search. Because the graph is constructed using weighted fused representations, its neighbor selection depends on the fusion weights rather than being independent of arbitrary query weights.

DEG~\cite{DEG} combines greedy Pareto frontier search for candidate acquisition with dynamic relative neighborhood graph (RNG) pruning to construct a weight-aware graph for two-component vector search. Its pruning mechanism assigns each edge an active weight interval, allowing queries to traverse edges according to their weights. However, this interval-based pruning mechanism relies on the one-dimensional weight space of two components and cannot be directly extended to an arbitrary number of components. Moreover, DEG acquires candidates by searching a partially constructed graph and updates its adjacency during incremental insertion, introducing construction dependencies and additional candidate-maintenance and edge-pruning costs.

A different approach organizes component-specific neighbor relationships into a joint graph, allowing queries to traverse connections from different components within a single search. HJG~\cite{HJG} constructs a hierarchical graph for each component and synchronizes the levels of corresponding nodes to support weighted search across component-specific neighbors. Subsequent work~\cite{HJG-Opt} also studies normalization and weight learning within this framework. However, HJG selects neighbors separately within each component, potentially retaining redundant edges in the joint graph, while hierarchical synchronization introduces dependencies among graph updates. In contrast, SJG uses component graphs to acquire candidates and performs weight-independent Skyline-based neighbor selection after merging them, allowing final adjacency construction to proceed independently across nodes.

Attribute-filtered ANNS~\cite{FilteredDiskANN,ACORN,NHQ,MA-NSW,DIGRA25} and vector-set retrieval~\cite{ColBERT,DESSERT} address related but distinct tasks. The former restricts search results through attribute predicates, while the latter evaluates similarities between sets of vectors. For example, MV-HNSW~\cite{MVHNSW26} indexes a set of token vectors as a single graph node. Our setting instead combines distances between corresponding components using per-query weights.

\begin{figure*}[t]
\centering
\includegraphics[width=\textwidth]{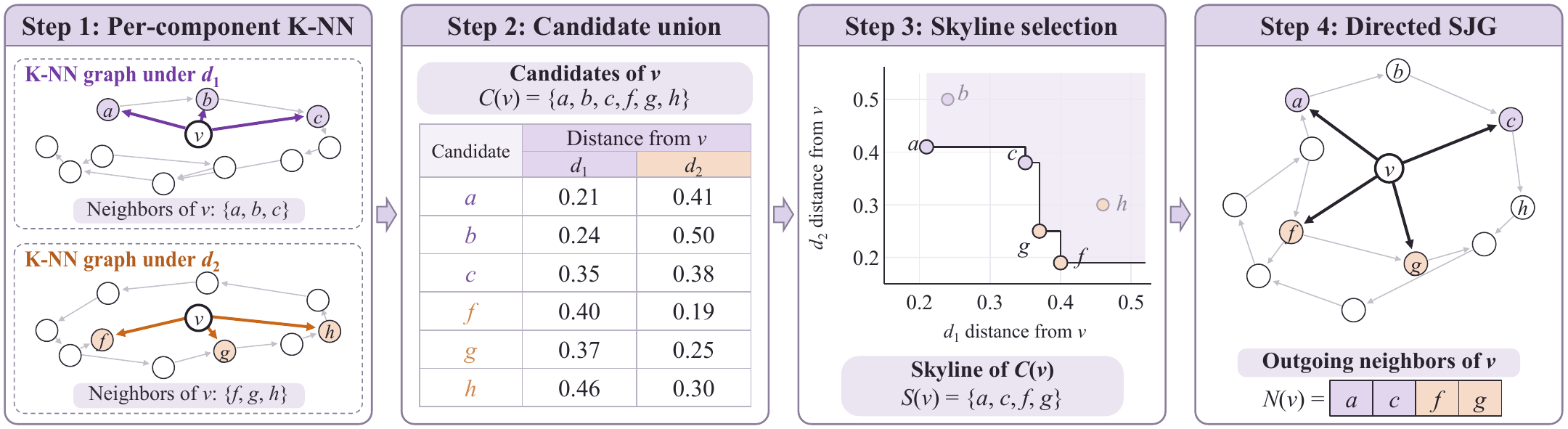}
\caption{SJG construction framework.}
\label{fig:construction}
\end{figure*}

\section{Problem Formulation}\label{sec:problem}

This section introduces the representation of multi-component data and the weighted distance used to compare objects, followed by a formal definition of the approximate nearest neighbor search problem considered in this paper. \hyperref[tab:notation]{Table~\ref*{tab:notation}} summarizes the main notation.

\begin{table}[!t]
\centering
\caption{Symbols and description}
\label{tab:notation}
\small
\renewcommand{\arraystretch}{1.14}
\begin{tabular}{@{}p{0.24\columnwidth}@{\hspace{0.025\columnwidth}}>{\raggedright\arraybackslash}p{0.735\columnwidth}@{}}
\hline
\textbf{Notation} & \textbf{Description} \\
\hline
$q,o$ & Query object, data object \\
$k,K$ & Number of search results returned per query, neighbor-list size per component graph \\
$w$ & Query weight vector \\
$d_i(\cdot,\cdot),D_w(\cdot,\cdot)$ & Normalized distance on component $i$, weighted distance under $w$ \\
$v,u$ & Graph node, candidate neighbor of $v$ \\
$L_i(v),A_i(v)$ & Out-neighbors of $v$, union of out- and in-neighbors of $v$ in component $i$ \\
$C(v)$ & Candidate set pooled across components for $v$ \\
$S(v)$ & Nondominated candidates in $C(v)$ \\
$\rho_i(u),B_v(u)$ & Rank of $u$ in component $i$, sum of $u$'s ranks across components \\
$N(v)$ & Final outgoing neighbors of $v$ \\
$\tau$ & Minimum out-degree \\
\hline
\end{tabular}
\end{table}

Let $O=\{o_1,o_2,\ldots,o_n\}$ be a dataset of $n$ objects, each consisting of $m$ components. Different components may have different data representations, such as numerical values, vectors, text, or sets, while each component maintains a consistent representation across objects. An object $o\in O$ is represented as the tuple in Eq.~\eqref{eq:object}.
\begin{equation}\label{eq:object}
o=(o^{(1)},o^{(2)},\ldots,o^{(m)}).
\end{equation}
Here, $o^{(i)}\in U_i$ denotes the value of object $o$ in component $i$, where $U_i$ is the corresponding data domain. Each component is associated with a distance function $d_i:U_i\times U_i\rightarrow\mathbb{R}_{\geq 0}$ that measures dissimilarity between component values. Different components may use different distance functions, such as $L_1$ distance, $L_2$ distance, edit distance, or Jaccard distance. Since component distances may have different numerical scales, we apply maximum-distance normalization by dividing each component distance by its corresponding maximum distance. We continue to use $d_i$ to denote the normalized component distance.

Given a query object $q=(q^{(1)},q^{(2)},\ldots,q^{(m)})$, a weight vector $w=(w_1,w_2,\ldots,w_m)$ specifies the relative importance of the components for the current query. A zero weight indicates that the corresponding component is excluded from distance evaluation. When all weight is assigned to a single component, the neighbor ordering depends only on that component. When multiple components have positive weights, their distances jointly determine the ordering. Let $I_w=\{i\mid w_i>0\}$ denote the set of active components. The weighted distance between $q$ and an object $o$ is defined in Eq.~\eqref{eq:distance}.
\begin{equation}\label{eq:distance}
D_w(q,o)=\sum_{i\in I_w}w_i\,d_i\bigl(q^{(i)},o^{(i)}\bigr).
\end{equation}
Here, $w_i\ge 0$ for each component $i$, and the weights are normalized so that $\sum_{i=1}^{m}w_i=1$.

\noindent\textbf{Multi-Component Weighted $k$NN Search.} Given a dataset $O$ and a query $(q,w,k)$, where $q$ is the query object, $w$ is the component weight vector, and $k$ is the number of requested neighbors, the objective is to retrieve the $k$ objects in $O$ with the smallest weighted distances to $q$. Let $R^*(q,w,k)$ denote the exact result set, where $|R^*|=k$ and $D_w(q,u)\leq D_w(q,v)$ for every $u\in R^*$ and $v\in O\setminus R^*$.

We study the approximate version of this problem. Given the same query, an approximate index returns a set $R$ of $k$ objects. Search quality is measured by $\mathrm{Recall@}k=|R\cap R^*|/k$. Approximate search aims to achieve high recall with limited search costs by reducing distance computations and graph traversal overhead.

\Needspace{5\baselineskip}
\section{The SJG Index}\label{sec:method}

A multi-component graph index must select the outgoing neighbors of every node before the query weights are known. A future query can use its weights to change how objects are evaluated and expanded, but it can discover further objects only through edges established during construction. Our design therefore pursues two related goals. Each node should have a neighbor list that serves queries with different weights, and final neighbor selection should proceed independently across nodes once candidate relationships have been established.

\subsection{SJG Framework}\label{sec:overview}

The Skyline Joint Graph (SJG) organizes neighbor relationships from different components into a static, single-layer directed graph $G=(O,E)$. Each object $v\in O$ is a node. The edge set $E$ contains a directed edge $(v,u)$ for each final outgoing neighbor $u\in N(v)$.

Combining neighbor relationships from different components does not, by itself, require synchronized hierarchies. We therefore examine the effect of the hierarchy on search by flattening a fully constructed HJG~\cite{HJG} while retaining its original neighbor relationships. The effect on search cost varies across datasets. Flattening reduces the cost on some datasets and increases it on others (\hyperref[sec:gains]{Section~\ref*{sec:gains}}), so the hierarchy does not provide a consistent search benefit. Maintaining synchronized levels, however, couples updates across component graphs and can trigger further updates to neighboring nodes. Taken together, these observations motivate a single-layer organization that retains cross-component connections without the cost of maintaining synchronized hierarchies.

With the single-layer structure in place, construction can be organized around two tasks. It must first obtain possible neighbor relationships and then decide which of them should become final outgoing edges. As illustrated in \hyperref[fig:construction]{Fig.~\ref*{fig:construction}}, candidate acquisition obtains approximate neighbor relationships separately from each component and pools them into a joint candidate set $C(v)$ for every node $v$. Joint neighbor selection then compares each candidate's distances across all components and determines the final adjacency using the Skyline as its basis. The key is to exploit component-wise dominance, which allows candidates to be compared without committing to a predefined weighting. In this way, the component graphs serve as sources of candidates, whereas queries traverse the single joint graph formed from the selected neighbors.

\subsection{Candidate Acquisition}\label{sec:candidate}

The single-layer organization removes the need to synchronize hierarchical levels, but a separate issue remains, namely how to obtain candidates. A single-layer graph could still be built incrementally. If objects are inserted one at a time by searching a partially constructed joint graph, the candidates available to a new node depend on the adjacency produced by earlier insertions. For a static dataset, however, candidate neighbors can be obtained independently from individual components without relying on a partially constructed joint graph. This allows us to acquire candidate relationships in advance and determine each node's final adjacency without depending on an evolving joint graph.

The neighbors of individual components can serve queries with different weight preferences. When all the weight is assigned to component $i$, the weighted neighbor ordering is determined entirely by $d_i$. The neighbors under each component therefore provide candidate connections relevant to queries that emphasize that component. To obtain these relationships, SJG uses NN-Descent~\cite{NN-Descent} to construct an approximate $K$-nearest-neighbor graph for each component, yielding a neighbor list $L_i(v)$ for object $v$. Here, $K$ controls the component neighbor-list size and differs from the number $k$ of results requested by a query. These component graphs can be constructed independently because each uses only its corresponding distance function $d_i$.

These neighbor lists, however, do not necessarily capture every relationship involving $v$. The $K$-nearest-neighbor relation need not be symmetric, even when the neighbors are computed exactly. An object $u$ may select $v$ as a neighbor without appearing in $v$'s own list. If we used only $L_i(v)$, such a relationship would be lost before the joint selection stage. SJG therefore forms the component-specific candidate set $A_i(v)$ in Eq.~\eqref{eq:component-candidates} from both the forward and reverse neighbors of $v$.
\begin{equation}\label{eq:component-candidates}
A_i(v)=L_i(v)\cup\{u\in O:v\in L_i(u)\}.
\end{equation}

SJG then takes the union across components. The resulting set $C(v)$ in Eq.~\eqref{eq:pool} contains each candidate only once, even if several components supply it.
\begin{equation}\label{eq:pool}
C(v)=\bigcup_{i=1}^{m}A_i(v).
\end{equation}

For example, the two components in \hyperref[fig:construction]{Fig.~\ref*{fig:construction}} contribute $\{a,b,c\}$ and $\{f,g,h\}$, giving $C(v)=\{a,b,c,f,g,h\}$. This union provides possible outgoing edges for node $v$ during construction. It serves a different role from merging the result lists of separate component searches. Once a query starts traversing the final graph, it can visit successive nodes and discover further objects through their neighborhoods.

At this point, the component graphs have supplied a pool of possible connections for each node. Yet this pool is only a starting point. When a query combines multiple components, being close in one component alone does not establish that a candidate should be retained in the joint graph. We must therefore compare the candidates in $C(v)$ across all components before determining the final adjacency.

\subsection{Skyline-Based Joint Neighbor Selection}\label{sec:selection}

Given the joint candidate set $C(v)$, the remaining task is to decide which candidates should become outgoing neighbors. Since the query weights are unknown, retaining every candidate may seem like a straightforward way to avoid excluding neighbors preferred by future queries. However, this leaves more neighbors to examine whenever a node is expanded. At the other extreme, selecting edges under a predefined weighted distance would favor one particular combination of components. We therefore need a selection rule that reduces unnecessary neighbors without fixing the query's weighting in advance.

\begin{figure}[t]
\centering
\includegraphics[width=\columnwidth]{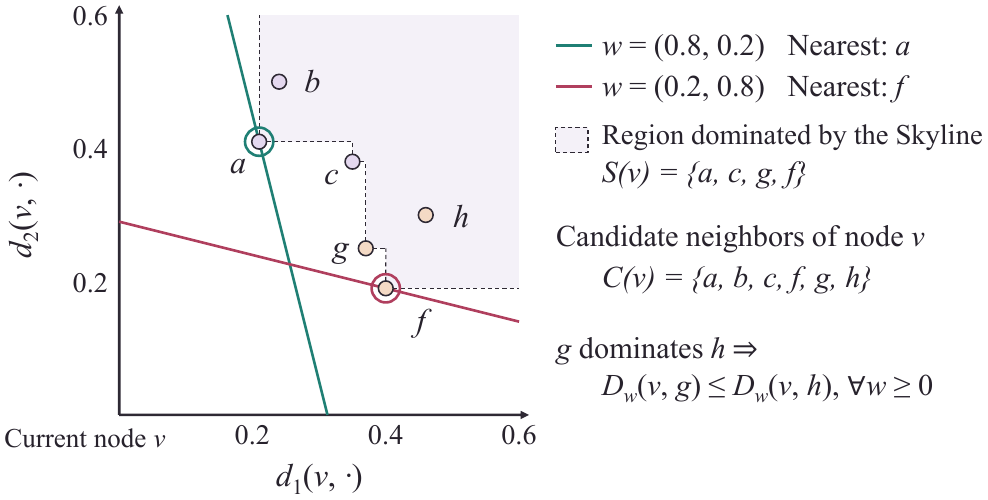}
\caption{Weight-dependent ordering and weight-independent dominance.}
\label{fig:candidate-dominance}
\end{figure}

Although the weights prevent us from fixing a complete ordering of the candidates, they do not make every comparison uncertain. \hyperref[fig:candidate-dominance]{Fig.~\ref*{fig:candidate-dominance}} illustrates the distinction. Changing the weights switches the nearest candidate from $a$ to $f$, so their relative preference cannot be determined in advance. By contrast, $g$ is no farther from $v$ than $h$ in every component. Its weighted distance to $v$ therefore cannot exceed that of $h$, regardless of how the nonnegative weights are assigned. Thus, while some comparisons vary with the weights, others remain valid for all admissible weightings. These weight-independent comparisons provide a basis for selecting neighbors before the query weights are known.

To compare two candidates, SJG examines their distances from $v$ in each component, writing $d_i(v,u)$ for the distance between $v^{(i)}$ and $u^{(i)}$.

\noindent\begin{minipage}{\columnwidth}
\begin{definition}[Dominance]\label{def:dominance} For two candidates $u,u'\in C(v)$, $u$ dominates $u'$, denoted by $u\prec_v u'$, if $u$ is no farther from $v$ in any component, as specified in Eq.~\eqref{eq:dominance}, and is strictly closer in at least one component.
\begin{equation}\label{eq:dominance}
d_i(v,u)\le d_i(v,u')\qquad\text{for every }i.
\end{equation}
\end{definition}
\end{minipage}\par\addvspace{7pt}

The definition captures exactly the weight-independent comparison illustrated above. It depends only on the component distances, not on the weights of a future query. SJG can therefore use this relation to identify candidates that are not dominated by any other candidate. They form the Skyline~\cite{Skyline01} of $v$, denoted by $S(v)$ in Eq.~\eqref{eq:skyline}.
\begin{equation}\label{eq:skyline}
S(v)=\{u\in C(v):\nexists u'\in C(v)\text{ such that }u'\prec_v u\}.
\end{equation}

Among the candidates in \hyperref[fig:construction]{Fig.~\ref*{fig:construction}}, $a$ dominates $b$, while $f$ and $g$ dominate $h$, giving $S(v)=\{a,c,f,g\}$. SJG uses this set as its initial neighbor set.

The dominance relation gives us a principled way to eliminate candidates. However, we must ensure that removing dominated candidates does not increase the minimum weighted distance from $v$ within $C(v)$ under any nonnegative query weights. The following proposition establishes this property.

\begin{proposition}[Weighted Minimum Preservation]\label{prop:cover} For any node $v$ with a nonempty finite candidate set $C(v)$ and any nonnegative weight vector $w$, restricting the candidates to $S(v)$ preserves the minimum weighted distance from $v$.
\begin{equation}\label{eq:min-preservation}
\min_{u\in S(v)}D_w(v,u)
=
\min_{u\in C(v)}D_w(v,u).
\end{equation}

\end{proposition}
\begin{proof} Consider a candidate $u'\in C(v)\setminus S(v)$. Since $u'$ is dominated, another candidate dominates it. Following the dominance relation in the finite set $C(v)$ eventually reaches a nondominated candidate $u\in S(v)$. By transitivity, $u\prec_v u'$. Since all weights are nonnegative, summing the component-wise inequalities preserves their direction, as shown in Eq.~\eqref{eq:weighted-order}.
\begin{equation}\label{eq:weighted-order}
\begin{aligned}
D_w(v,u) &= \sum_{i=1}^{m}w_i d_i(v,u) \\
         &\le \sum_{i=1}^{m}w_i d_i(v,u') = D_w(v,u').
\end{aligned}
\end{equation}

Thus, every excluded candidate has a Skyline candidate with no greater weighted distance to $v$. Since $S(v)\subseteq C(v)$, the two sets have the same minimum weighted distance. \end{proof}

SJG therefore need not decide during construction which candidate will minimize the weighted distance from $v$ under a future weighting. Regardless of the nonnegative weighting eventually specified, the Skyline contains at least one candidate that attains the minimum weighted distance from $v$ in the original candidate set.

\noindent\textbf{Minimum Out-Degree.} Although the Skyline preserves the minimum weighted distance within the candidate set, it can contain very few neighbors. For example, if one candidate dominates all others, the Skyline contains only that candidate. To avoid excessively small outgoing neighborhoods, SJG retains all Skyline candidates and supplements them only when their number is below $\min(\tau,|C(v)|)$, where $\tau$ is the minimum out-degree.

When supplementation is needed, SJG determines the order of additional candidates using their ranks within each component. Let $\rho_i(u)$ denote the rank of candidate $u$ among the candidates in $C(v)$ under component distance $d_i(v,u)$, with smaller ranks assigned to closer candidates. The rank sum $B_v(u)$ in Eq.~\eqref{eq:rank-sum} determines which candidates are added first.
\begin{equation}\label{eq:rank-sum}
B_v(u)=\sum_{i=1}^{m}\rho_i(u).
\end{equation}

When supplementation is needed, SJG adds candidates from $C(v)\setminus S(v)$ in ascending order of $B_v(u)$ until $|N(v)|=\min(\tau,|C(v)|)$. The final neighbor set $N(v)$ contains all Skyline neighbors and any candidates added during supplementation.

Every final neighbor comes from $C(v)$, and every Skyline neighbor is retained, as expressed in Eq.~\eqref{eq:inclusion}. The resulting out-degree in Eq.~\eqref{eq:degree} equals the Skyline size unless additional candidates are needed and available to reach the minimum out-degree $\tau$.
\begin{gather}
S(v)\subseteq N(v)\subseteq C(v),\label{eq:inclusion}\\
|N(v)|=\max\bigl(|S(v)|,\min(\tau,|C(v)|)\bigr).\label{eq:degree}
\end{gather}

It follows from \hyperref[prop:cover]{Proposition~\ref*{prop:cover}} and the inclusions in Eq.~\eqref{eq:inclusion} that the final neighbor set attains the same minimum weighted distance as the full candidate set for every nonnegative weighting.

Thus, the Skyline determines the core neighbors of each node, while minimum-degree supplementation adds extra connections only when necessary. Together, they determine the final adjacency.

\noindent\textbf{Scale Invariance.} The construction steps described above, including candidate acquisition, dominance testing, and rank-based supplementation, rely only on distance orderings within individual components. A strictly increasing transformation of the distances in any component therefore preserves the comparisons used by these steps. Provided that the randomized choices, processing order, and tie-handling rules remain consistent, the constructed adjacency is unchanged. This invariance applies to graph construction rather than query rankings. During search, weighted distances are evaluated using the specified distance functions and weights, so the ranking under fixed weights may still change when the distance values change.

\subsection{Parallel Graph Construction and Complexity}\label{sec:algorithm}

In the following, we discuss how SJG supports parallel graph construction and analyze its time and space complexity.

\noindent\textbf{Parallel Construction.} Candidate acquisition can proceed independently across components because each uses only its own distance function. Once the forward and reverse relationships $A_i(v)$ are established, final neighbor selection also becomes independent across nodes. Each node $v$ reads the fixed candidate relationships and object data, computes its candidate-distance vectors, and writes only its own adjacency $N(v)$, without reading or modifying the final adjacency of other nodes. Consequently, nodes can select their neighbors in parallel without searching an evolving joint graph or maintaining synchronized hierarchies.

\hyperref[alg:build]{Algorithm~\ref*{alg:build}} summarizes the two stages. Lines 1--9 establish component-specific candidate relationships, while lines 10--20 independently determine each node's final adjacency. The algorithm then selects $o_1$ as the fixed entry node $v_0$.

\begin{algorithm}[t]
\caption{SJG Construction}\label{alg:build}
\begin{algorithmic}[1]
\Require $O$, distances $d_1,\ldots,d_m$, $K$, minimum out-degree $\tau$
\Ensure Joint graph $G$ and entry node $v_0$
\For{$i=1,\ldots,m$}
  \State Initialize $A_i(v)\gets\emptyset$ for every $v\in O$
  \State $L_i\gets \mathrm{NNDescent}(O,d_i,K)$
  \For{each $v\in O$}
    \For{each $u\in L_i(v)$}
      \State Add $u$ to $A_i(v)$; add $v$ to $A_i(u)$
    \EndFor
  \EndFor
\EndFor
\For{each $v\in O$ \textbf{in parallel}}
  \State $C(v)\gets$ deduplicated union of $A_i(v)$
  \State Compute and store $d_i(v,u)$ for each $u\in C(v)$ and each component $i$
  \State $S(v)\gets$ nondominated candidates in $C(v)$
  \State $N(v)\gets S(v)$
  \If{$|N(v)|<\min(\tau,|C(v)|)$}
    \State \parbox[t]{\dimexpr\linewidth-\algorithmicindent*2\relax}{\raggedright Compute $B_v(u)$ from the component-wise ranks in $C(v)$}
    \State \parbox[t]{\dimexpr\linewidth-\algorithmicindent*2\relax}{\raggedright Add $u\in C(v)\setminus S(v)$ to $N(v)$ in ascending order of $B_v(u)$ until $|N(v)|=\min(\tau,|C(v)|)$}
  \EndIf
  \State $G[v]\gets N(v)$
\EndFor
\State $v_0\gets o_1$
\State \Return $G$, $v_0$
\end{algorithmic}
\end{algorithm}

\Needspace{4\baselineskip}
\noindent\textbf{Time Complexity.} Let $n=|O|$, and let $T_i(n,K)$ denote the work required to construct the approximate $K$-nearest-neighbor graph for component $i$. Building all $m$ component graphs therefore requires $\sum_{i=1}^{m}T_i(n,K)$ total work.

Each component contributes at most $nK$ forward and $nK$ reverse entries. Hash-based deduplication of these entries takes $O(nmK)$ expected work. Let $c_v=|C(v)|$ denote the number of distinct candidates at node $v$. Deduplication can only reduce the number of entries, so the total candidate count satisfies $\sum_{v\in O}c_v\le 2nmK$.

During joint neighbor selection, component distances are computed once and reused for dominance testing and ranking. For each node $v$, constructing candidate-distance \mbox{vectors} requires at most $mc_v$ component-distance evaluations, while pairwise Skyline selection requires $O(mc_v^2)$ scalar comparisons without additional distance evaluations. If supplementation is needed, computing component-wise ranks costs $O(mc_v\log c_v)$, summing the ranks costs $O(mc_v)$, and sorting the remaining candidates costs $O(c_v\log c_v)$ comparisons.

Across all nodes, candidate-distance vectors require at most $m\sum_{v\in O}c_v$ component-distance evaluations. With unit-cost component-distance evaluations, the expected total construction work is \mbox{$O(\sum_{i=1}^{m} T_i(n,K) + nmK + m\sum_{v\in O} c_v^2)$}.

\noindent\textbf{Space Complexity.} Let $\bar r$ denote the average final out-degree. The final graph stores one adjacency list per object, requiring $O(n\bar r)$ space. Forward and reverse candidate relationships require $O(nmK)$ auxiliary space, while processing node $v$ requires $O(mc_v)$ temporary space for candidate-distance vectors. This temporary storage can be released after processing each node, although concurrent workers require separate storage.

Since SJG retains the entire Skyline and uses $\tau$ only for minimum-degree supplementation, the final out-degree depends on the candidate set and its Skyline rather than a fixed maximum-degree bound.

\subsection{Query Processing}\label{sec:search}

\begin{algorithm}[!t]
\caption{SJG Search}\label{alg:search}
\begin{algorithmic}[1]
\Require $G$, entry node $v_0$, query $(q,w,k)$, $\searchbreadth\ge k$
\Ensure Approximate top-$k$ objects
\State $H\gets$ empty min-heap; $W\gets$ empty max-heap
\State $t\gets D_w(q,v_0)$
\State Insert $(t,v_0)$ into $H$ and $W$; $V\gets\{v_0\}$
\While{$H$ is not empty}
  \State $(t_u,u)\gets H.\mathrm{popMin}()$
  \If{$|W|=\searchbreadth$ and $t_u>\mathrm{maxDistance}(W)$}
    \State \textbf{break}
  \EndIf
  \For{each $x\in G[u]$}
    \State \textbf{if} $x\in V$ \textbf{then continue}
    \State Add $x$ to $V$
    \State $t_x\gets D_w(q,x)$
    \If{$|W|<\searchbreadth$ or $t_x<\mathrm{maxDistance}(W)$}
      \State Insert $(t_x,x)$ into $H$ and $W$
      \State Retain at most $\searchbreadth$ closest objects in $W$
    \EndIf
  \EndFor
\EndWhile
\State \Return the $k$ closest objects in $W$
\end{algorithmic}
\end{algorithm}

Once construction is complete, the graph adjacency remains fixed, while query weights determine how encountered objects are evaluated and prioritized for expansion. Given a query $(q,w,k)$, SJG applies the weighted distance defined in Eq.~\eqref{eq:distance} to each encountered object $u$. Only components with positive weights require a distance evaluation.

SJG uses best-first search (\hyperref[alg:search]{Algorithm~\ref*{alg:search}}) to explore promising neighbors while retaining alternative directions of expansion. A min-heap $H$ manages candidates awaiting expansion, while a max-heap $W$ retains at most $\searchbreadth$ potential results and determines the distance threshold for admitting new candidates. The parameter $\searchbreadth\ge k$ controls search breadth, and a visited set $V$ avoids repeated evaluations. Search begins from the fixed entry node $v_0$.

During traversal, the closest candidate in $H$ is expanded, and its unvisited neighbors are evaluated using $D_w$. A neighbor is admitted to both heaps if $W$ is not full or its weighted distance is smaller than the farthest distance in $W$. Importantly, removing an object from $W$ does not remove it from $H$, so it may still be expanded.

Search terminates when $H$ is empty or the closest candidate awaiting expansion is farther than the farthest object in a full $W$. The $k$ closest objects in $W$ are returned. Throughout this process, query weights affect candidate evaluation and expansion order, while the graph adjacency remains unchanged.

\section{Experiments}\label{sec:exp}

We first compare SJG with existing methods under different query weights. We then examine how the hierarchy, candidate sources, and neighbor selection affect search performance and how sensitive SJG is to its construction parameters and the number of requested results. Finally, we compare construction time and index size across methods and evaluate query performance and construction cost on larger datasets.

\subsection{Experimental Setup}\label{sec:setup}

\textbf{Datasets.} We evaluate SJG on six real-world datasets, with their statistics summarized in \hyperref[tab:datasets]{Table~\ref*{tab:datasets}}. Corel~\cite{UCI-Corel} contains four visual features, including a 32-dimensional color histogram, a 32-dimensional color layout histogram, 9-dimensional color moments, and 16-dimensional co-occurrence texture. NUS-WIDE~\cite{NUS-WIDE} contains five visual features, including a 64-dimensional color histogram, a 144-dimensional color correlogram, a 73-dimensional edge direction histogram, 128-dimensional wavelet texture, and 225-dimensional block-wise color moments.

For Tiny ImageNet~\cite{TinyImageNet}, we use the training images and extract five features: a 256-dimensional HSV color histogram, 9-dimensional color moments, 32-dimensional Gabor texture, 324-dimensional HOG, and 2048-dimensional ResNet-50~\cite{ResNet} representations. For COCO~\cite{COCO}, we use images and captions from the 2017 training set. Its eight components include four visual features, CLIP~\cite{CLIP} and ResNet-50 image representations, and MPNet~\cite{MPNet} and CLIP text representations. Multiple captions of each image are encoded and aggregated into one text representation. The deep representations in Tiny ImageNet and COCO are \(L_2\)-normalized.

Food and Airbnb contain heterogeneous components beyond vectors. Food is derived from Open Food Facts~\cite{OFF} and contains 40,000 objects, each with five nutritional values per 100 grams, a food category, and a 768-dimensional MPNet representation of its product name. Airbnb uses listing snapshots from 119 locations in Inside Airbnb~\cite{InsideAirbnb}, dated between September 11, 2025, and June 19, 2026. Each listing contains its location, name, price, guest capacity, and a 64-dimensional multi-hot vector over the 64 most frequent amenities.

We compute the distances between product-name vectors in Food and between amenity vectors in Airbnb using the word cosine distance (WCD), as defined in Eq.~\eqref{eq:wcd}~\cite{HJG}.
\begin{equation}\label{eq:wcd}
d_{\mathrm{WCD}}(x,y)=
\arccos\left(
\frac{x^\top y}{\|x\|_2\|y\|_2}
\right).
\end{equation}

\begin{table}[t]
\centering
\caption{Dataset statistics.}\label{tab:datasets}
\small
\renewcommand{\arraystretch}{1.12}
\setlength{\tabcolsep}{3pt}
\begin{tabular}{@{}>{\centering\arraybackslash}m{1.47cm}>{\centering\arraybackslash}m{1.37cm}>{\centering\arraybackslash}m{0.31cm}>{\raggedright\arraybackslash}m{5.05cm}@{}}
\toprule
Dataset & $n$ & $m$ & Distance functions \\
\midrule
Corel & 65,615 & 4 & $L_1$: color histogram, color layout\newline $L_2$: color moments, co-occurrence texture \\
\midrule
NUS-WIDE & 268,648 & 5 & $L_1$: color histogram, color correlogram, edge\newline $L_2$: wavelet texture, block-wise color moments \\
\midrule
Tiny ImageNet & 100,000 & 5 & $L_1$: HSV color histogram\newline $L_2$: color moments, Gabor, HOG, ResNet-50 \\
\midrule
COCO & 117,287 & 8 & $L_1$: HSV color histogram, LBP texture, edge\newline $L_2$: block-wise color moments, ResNet-50, CLIP image and text, MPNet text \\
\midrule
Food & 40,000 & 7 & $L_1$: energy, fat, sugar, salt, protein\newline Edit distance: food category\newline Word cosine distance: product name \\
\midrule
Airbnb & 1,135,146 & 5 & $L_1$: price, guest capacity\newline $L_2$: latitude and longitude\newline Edit distance: listing name\newline Word cosine distance: amenities \\
\bottomrule
\end{tabular}
\end{table}

\noindent\textbf{Baselines.} We compare SJG with three composite methods, C-HNSW, C-NSG, and C-Vamana, and the joint graph HJG~\cite{HJG,HJG-code}. The composite methods build an HNSW, NSG, or Vamana index for each component using public implementations~\cite{hnswlib-code,NSG-code,ParlayANN-code}. We implement the combination, which retrieves \(\searchbreadth\) candidates from each component, merges and deduplicates them, and reranks them by weighted distance.

\noindent\textbf{Metrics.} We measure result quality using Recall@\(k\), with \(k=10\) by default. Ground truth is computed by linear scan. We measure query efficiency by queries per second (QPS) or mean query latency. For the ablation study, we count each evaluation of a component distance function as one distance computation.

\begin{figure*}[t]
\centering
\includegraphics[width=0.810461316\textwidth]{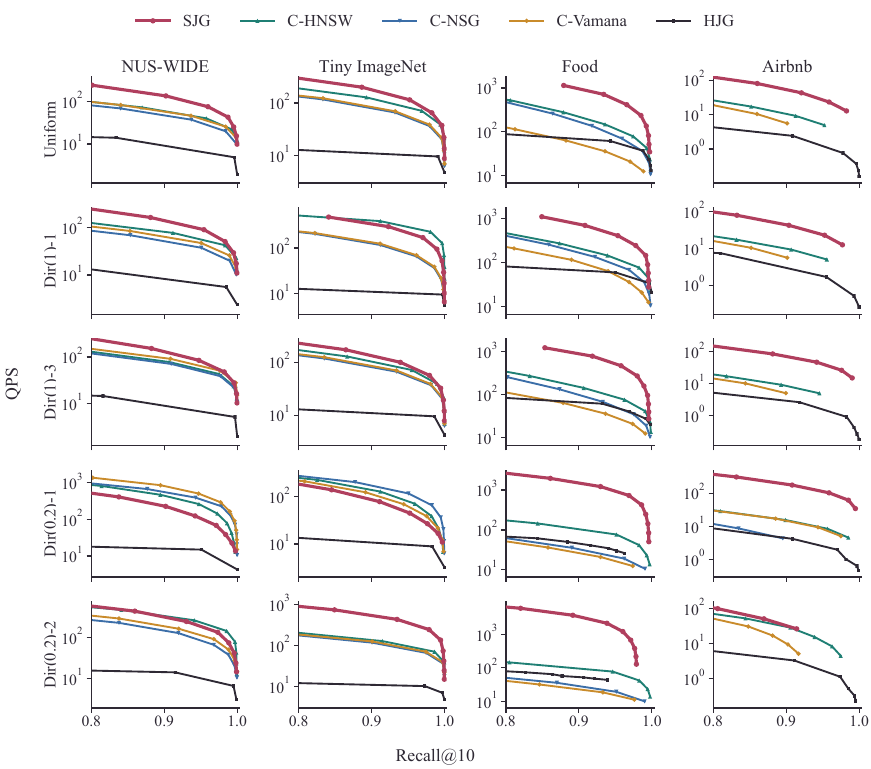}
\caption{Query performance under different weight settings.}
\label{fig:weights}
\end{figure*}

\noindent\textbf{Protocol.} We use 1,000 queries per dataset. \hyperref[sec:weights]{Section~\ref*{sec:weights}} evaluates both uniform and random weights, while the remaining experiments use uniform weights. SJG's search parameter \(\searchbreadth\) and the number of candidates returned by each component index in the composite methods are both doubled from 16 to 4096. In the parameter study, we test only search budgets with \(\searchbreadth\geq k\). For HJG, we evaluate the distance threshold factor \(\varepsilon\) at 0.75, 1.0, 1.1, 1.2, 1.3, 1.4, and 1.5. For \hyperref[fig:weights]{Fig.~\ref*{fig:weights}}, we show the best recall--QPS trade-offs across the tested search configurations. For comparisons across methods at Recall@10 targets of 0.90 and 0.95, we report the highest QPS among configurations reaching each target. Methods that fail to reach a target are excluded from its comparison.

All queries run on a single thread. Construction uses 12 threads for SJG and the composite methods, while HJG runs on a single thread. We additionally report single-thread construction time for SJG in \hyperref[sec:build]{Section~\ref*{sec:build}}. For index construction, SJG uses \(K=15\) and \(\tau=16\) by default. We vary these parameters in \hyperref[sec:sensitivity]{Section~\ref*{sec:sensitivity}}. C-HNSW uses \(M=16\) and \(\mathit{efConstruction}=200\); C-NSG uses 100 initial neighbors, \(R=32\), \(L=40\), and \(C=500\); and C-Vamana uses \(R=32\), \(L=200\), \(\alpha=1.2\), and two construction passes. HJG uses \(\mathrm{NN}_{\mathrm{num}}=15\) and \(\mathrm{Max}_{n}=1500\).

All experiments run on a server with two Intel Xeon E5-2680 v3 processors (2.50~GHz, 24 physical cores) and 504~GiB of memory under Ubuntu 20.04. The source code for SJG is available at \url{https://github.com/Ma-raca/SJG-code}.

\Needspace{8\baselineskip}
\subsection{Query Performance under Different Weights}\label{sec:weights}

\begingroup\setlength{\parskip}{0pt}
To evaluate how SJG performs under varying query weights, we consider both equal and randomly sampled weight assignments. Under equal weighting, each of the $m$ components receives weight $1/m$. We also sample three weight vectors per dataset from each of two symmetric Dirichlet distributions, giving seven weight settings per dataset and 42 settings across six datasets. All seven settings on each dataset use the same prebuilt index. With $\alpha=1$, sampling is uniform over all nonnegative weight vectors whose entries sum to one. With $\alpha=0.2$, sampled vectors are more likely to concentrate most of the weight on a few components. We denote the \mbox{$j$-th} sampled weight vector by $\mathrm{Dir}(\alpha)\text{-}j$. Due to space limitations, \hyperref[fig:weights]{Fig.~\ref*{fig:weights}} presents five weight settings on four datasets.

At a Recall@10 target of 0.90, SJG outperforms the fastest composite baseline in 36 of the 42 settings, achieving a median speedup of 2.05$\times$ and a maximum speedup of 27.9$\times$. Against HJG, SJG is faster in all 42 settings, with a median speedup of 21.4$\times$. Under equal weights, SJG is the fastest on all six datasets. It is 1.6--4.8$\times$ faster than the fastest composite method and 12--28$\times$ faster than HJG. The conclusion holds at a higher recall target. At Recall@10 = 0.95, SJG is faster in 34 of the 40 comparable settings. Under equal weights, the smallest gap is on COCO, where SJG and C-HNSW achieve 89.0 and 83.9 QPS.

Moreover, the gap to the composite methods is largest on the datasets with non-vector components. In the 14 settings of Food and Airbnb, the speedup of SJG is at least 4.5$\times$ in all but one, and it reaches 27.9$\times$. On the four vector-only datasets, the median speedup is 1.59$\times$. On Food, the composite methods require at least 256 returned candidates per component to reach Recall@10 = 0.90 under every weight setting, whereas SJG reaches the same target with a search breadth of 32--128. On Airbnb under uniform weights, C-HNSW needs $\searchbreadth=2048$ and C-Vamana needs 4096, and C-NSG reaches a recall of only 0.58 at 4096. SJG needs only 1024.

In contrast, the settings where SJG falls behind concentrate on weights dominated by a single component. Among the 12 settings whose largest component weight is about 0.8 or higher, SJG falls behind in 4. In the other 30 settings, the largest component weight is at most 0.71, and SJG falls behind in only 2. This effect is most pronounced on NUS-WIDE under Dir(0.2)-1, where the ranking is driven primarily by similarity in block-wise color moments. Candidates retrieved from that component alone can therefore be competitive under the weighted distance. At the target recall, C-Vamana achieves 504.3 QPS, compared with 225.9 QPS for SJG.

\par\endgroup

\subsection{Ablation Study}\label{sec:gains}

\begin{figure}[!t]
\centering
\includegraphics[width=\columnwidth]{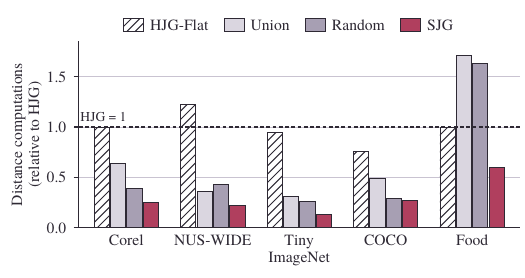}
\caption{Ablation study on distance computations.}
\label{fig:ablation}
\end{figure}

We conduct ablation studies to examine the effects of three key design choices: the hierarchy, candidate acquisition, and neighbor selection. We compare the minimum number of distance computations required to reach Recall@10 = 0.90 under uniform weights (\hyperref[fig:ablation]{Fig.~\ref*{fig:ablation}}). HJG-Flat removes the hierarchy from HJG while preserving its original neighbor relationships. Union constructs a single-layer graph by combining the forward and reverse neighbors from the component-specific \(K\)-NN graphs, retaining all candidates as outgoing neighbors. SJG retains the Skyline of each candidate set and adds further candidates only when needed to meet the minimum out-degree. For each node, Random retains the same number of neighbors as SJG but selects them randomly from the Union candidates.

\begin{figure*}[!t]
\centering
\includegraphics[width=\textwidth]{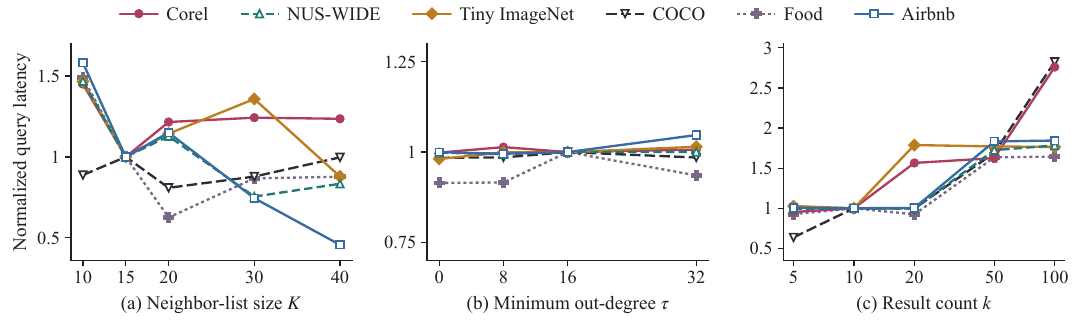}
\caption{Parameter sensitivity of SJG.}
\label{fig:sensitivity}
\end{figure*}

Skyline selection reduces distance computations on the five datasets. Compared with Union, SJG reduces the number of distance computations by 41\%--65\%. HJG needs 1.7--7.4$\times$ as many distance computations as SJG.

Notably, the effects of the hierarchy and the candidate source vary across datasets. Removing the hierarchy reduces the number of distance computations by 24\% on COCO but increases it by 23\% on NUS-WIDE. On Corel and Food, HJG has only one level, so HJG-Flat is identical to HJG. Compared with HJG-Flat, using the union of the per-component neighbors as candidates reduces the number of distance computations by 36\%--71\% on four datasets but increases it by 72\% on Food.

In particular, the benefit of joint neighbor selection comes from which neighbors SJG keeps, not only from how many. Random keeps as many neighbors as SJG but does not guarantee the weighted-minimum preservation property of \hyperref[prop:cover]{Proposition~\ref*{prop:cover}}. Except on COCO (1.09$\times$), Random needs 1.60--2.75$\times$ as many distance computations as SJG. On NUS-WIDE, it needs even more than Union, which keeps all candidates.

In summary, Skyline selection is what consistently reduces distance computations on the five datasets, while the effects of the hierarchy and the candidate source vary with the dataset.

\subsection{Parameter Sensitivity}\label{sec:sensitivity}

We examine the sensitivity of SJG to the per-component neighbor-list size \(K\), minimum out-degree \(\tau\), and result count \(k\) on six datasets under uniform weights. We vary one parameter at a time, keeping the others at their default values (\(K=15\), \(\tau=16\), and \(k=10\)). For each configuration, we report the minimum measured query latency achieving \(\mathrm{Recall@}k \geq 0.95\) by varying the search parameter \(\searchbreadth\) (\hyperref[fig:sensitivity]{Fig.~\ref*{fig:sensitivity}}). Latency is normalized to that of the default configuration on each dataset.

Increasing \(K\) can substantially improve search efficiency. On Airbnb, increasing \(K\) from 10 to 40 reduces query latency from 68.4 to 19.7~ms, achieving a \(3.47\times\) speedup. Meanwhile, the default \(K=15\) uses 33.2\%--47.9\% fewer edges than \(K=40\) across the six datasets, providing a more compact graph. SJG is less sensitive to \(\tau\): the maximum latency variation across the tested values is below 9.4\% on every dataset, while \(\tau=16\) adds less than 2.9\% more edges than \(\tau=0\). Retrieving 100 rather than 10 results requires only 1.64--2.83\(\times\) the query time across all six datasets, without rebuilding the index.

\begin{table}[!t]
\centering
\caption{Construction time (minutes).}\label{tab:buildtime}
\footnotesize
\setlength{\tabcolsep}{3.5pt}
\begin{tabular}{@{}lrr@{\hspace{10pt}}rrrr@{}}
\toprule
\multirow{2}{*}{Dataset} & \multicolumn{2}{c}{1 thread} & \multicolumn{4}{c}{12 threads} \\
\cmidrule(lr){2-3}\cmidrule(l){4-7}
 & \multicolumn{1}{c}{HJG} & \multicolumn{1}{c}{SJG} & \multicolumn{1}{c}{SJG} & \multicolumn{1}{c}{C-HNSW} & \multicolumn{1}{c}{C-NSG} & \multicolumn{1}{c}{C-Vamana} \\
\midrule
Corel & 2.81 & 0.77 & 0.18 & 0.06 & 0.56 & 0.21 \\
NUS-WIDE & 141.08 & 11.55 & 2.54 & 1.88 & 18.54 & 4.56 \\
Tiny ImageNet & 48.89 & 10.65 & 1.61 & 1.31 & 23.76 & 7.33 \\
COCO & 62.68 & 20.50 & 3.31 & 1.96 & 33.71 & 10.97 \\
Food & 5.25 & 1.75 & 0.35 & 0.27 & 6.87 & 1.21 \\
Airbnb & 7,816.03 & 38.91 & 8.20 & 22.77 & 73.23 & 105.90 \\
\bottomrule
\end{tabular}
\end{table}

\begin{table}[t]
\centering
\caption{Graph index size (\textnormal{MiB}).}\label{tab:size}
\footnotesize
\setlength{\tabcolsep}{4pt}
\begin{tabular}{@{}lrrrrr@{}}
\toprule
Dataset & \multicolumn{1}{c}{SJG} & \multicolumn{1}{c}{HJG} & \multicolumn{1}{c}{C-HNSW} & \multicolumn{1}{c}{C-NSG} & \multicolumn{1}{c}{C-Vamana} \\
\midrule
Corel & 8.16 & 34.26 & 36.59 & 13.62 & 19.92 \\
NUS-WIDE & 62.61 & 178.08 & 189.54 & 103.22 & 131.33 \\
Tiny ImageNet & 21.46 & 65.17 & 69.37 & 34.65 & 46.48 \\
COCO & 48.30 & 121.99 & 131.71 & 75.47 & 99.34 \\
Food & 9.50 & 21.55 & 38.79 & 25.87 & 16.42 \\
Airbnb & 165.08 & 732.79 & 807.78 & 490.25 & 407.02 \\
\bottomrule
\end{tabular}
\end{table}

\subsection{Construction Cost and Index Size}\label{sec:build}

SJG builds faster than HJG, and the gap is largest on the two largest datasets. On a single thread, HJG takes 3.0--4.6$\times$ as long as SJG on Corel, Tiny ImageNet, COCO, and Food, and 12.2$\times$ as long on NUS-WIDE. On Airbnb, HJG takes about 130 hours, whereas SJG takes only 39 minutes, a 201$\times$ difference (\hyperref[tab:buildtime]{Table~\ref*{tab:buildtime}}).

Moreover, SJG construction can be parallelized. With 12 threads, SJG builds 4.3--6.6$\times$ faster than on a single thread. On Airbnb, the construction time of SJG drops to 8.2 minutes, less than that of every composite method. On the smaller datasets, C-HNSW builds fastest.

In addition, SJG has the smallest graph index. On the six datasets, the graph index of SJG is 36.0\%--59.4\% smaller than that of the smallest baseline. The graph index of HJG is 2.3--4.4$\times$ as large as that of SJG (\hyperref[tab:size]{Table~\ref*{tab:size}}). This agrees with the degrees of the methods. The average degree of SJG is 31.1--106.9 and is highest on COCO, which has the most components ($m=8$). The sum of the average degrees of the per-component graphs is 51.3--215.9 for the composite methods and 113.9--240.0 for HJG. SJG stores only one single-layer graph, whereas the composite methods store one graph per component, and HJG also stores adjacency by level.

\begin{figure}[!t]
\centering
\includegraphics[width=\columnwidth]{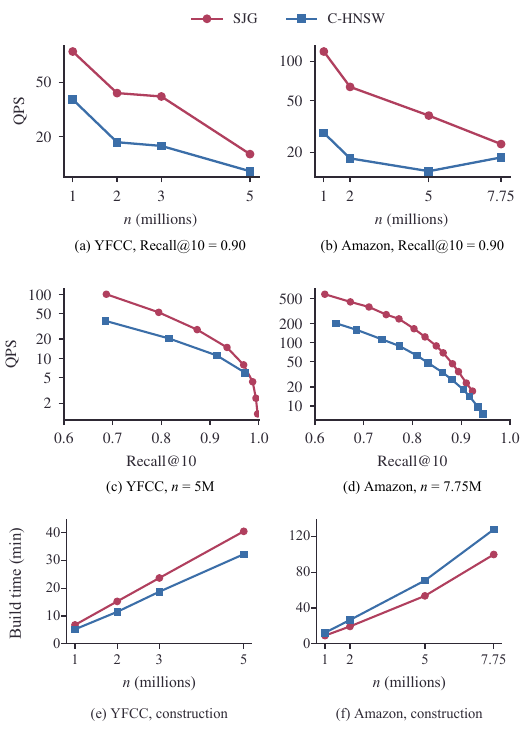}
\caption{Scalability on YFCC and Amazon.}
\label{fig:scale}
\end{figure}

\subsection{Scalability}\label{sec:scale}

We examine how SJG behaves on two larger datasets as their size grows. YFCC uses the LIRE~\cite{LIRE} features of YFCC100M~\cite{YFCC100M} images released in Multimedia Commons. It has four components, a 144-dimensional CEDD, a 120-dimensional color layout, an 80-dimensional edge histogram, and a 60-dimensional Gabor texture, compared by $L_1$, $L_2$, $L_1$, and $L_2$ distance, respectively. Amazon uses the product metadata of Amazon Reviews 2023~\cite{AmazonReviews23}. It has five components, the store name, the title, the category set, the price, and the rating. The store name is compared by edit distance, the 384-dimensional MiniLM~\cite{MiniLM} representation of the title by WCD, the category set by Jaccard distance, and the price and the rating by $L_1$ distance. On each dataset, we take nested prefix subsets, with 1M, 2M, 3M, and 5M objects for YFCC and 1M, 2M, 5M, and 7.75M objects for Amazon. A separate set of 1,000 objects serves as the queries. We compare SJG with C-HNSW under uniform weights. C-HNSW is the fastest composite method in 34 of the 42 settings in \hyperref[sec:weights]{Section~\ref*{sec:weights}}. Construction uses 12 threads, and queries run on a single thread.

\hyperref[fig:scale]{Figs.~\ref*{fig:scale}(a)} and \hyperref[fig:scale]{\ref*{fig:scale}(b)} show that SJG has higher query throughput than C-HNSW at every scale, although its lead narrows as the datasets grow. At Recall@10 = 0.90, SJG is 2.3$\times$ faster than C-HNSW on YFCC at 1M--3M and 1.3$\times$ faster at 5M. On Amazon, the speedup decreases from 4.2$\times$ at 1M to 1.3$\times$ at 7.75M. At the largest scale, \hyperref[fig:scale]{Figs.~\ref*{fig:scale}(c)} and \hyperref[fig:scale]{\ref*{fig:scale}(d)} show that SJG maintains higher QPS than C-HNSW for recall between 0.6 and 0.9.

Notably, on Amazon, the highest recall that SJG reaches decreases as the dataset grows. With $\searchbreadth=4096$, the recall of SJG decreases from 0.975 at 1M to 0.923 at 7.75M, and that of C-HNSW decreases from 0.960 to 0.945. At 5M and above, SJG does not reach Recall@10 = 0.95 within this budget. On YFCC, both methods reach a recall above 0.97 at every scale.

\hyperref[fig:scale]{Figs.~\ref*{fig:scale}(e)} and \hyperref[fig:scale]{\ref*{fig:scale}(f)} show how construction time grows with dataset size. On YFCC, C-HNSW builds faster at all four scales. As the dataset grows from 1M to 5M objects, its construction time rises from 5.2 to 32.2 minutes, while SJG's rises from 6.8 to 40.5 minutes. On Amazon, SJG builds faster at every scale, and its lead grows from about 3 minutes at 1M objects to 28 minutes at 7.75M.

\section{Conclusion}\label{sec:conclusion}

In this paper, we present the Skyline Joint Graph (SJG), a graph index for multi-component approximate nearest neighbor search with per-query weights. To address the challenge of selecting neighbors before query weights are known, SJG pools component-specific candidates and applies weight-independent Skyline pruning to construct a compact joint graph. By separating candidate acquisition from joint neighbor selection, SJG also enables parallel graph construction without hierarchical synchronization or cascading updates. Extensive experiments on six real-world datasets demonstrate the effectiveness of SJG. At Recall@10 = 0.90, SJG outperforms the fastest composite baseline in 36 of 42 dataset--weight combinations, achieving a median speedup of 2.05$\times$. Its graph index is at least 36\% smaller than the smallest baseline index, while its single-threaded construction is up to 201$\times$ faster than HJG. In future work, we will investigate how to preserve efficient graph navigation while reducing the number of retained Skyline neighbors.

\section*{Acknowledgment}

This work is partially supported by the NSFC project 62532007.

\section*{AI-Generated Content Acknowledgment}

We used generative AI tools to edit selected sentences for grammar, clarity, and readability.

\bibliographystyle{IEEEtran}
\bibliography{refs}

\end{document}